# A HIGH RESOLUTION MAP OF $^{12}$CO J = 6-5 EMISSION IN THE STARBURST GALAXY M82

E. R. SEAQUIST AND S.W. LEE

Department of Astronomy & Astrophysics, University of Toronto, Toronto, ON M5S 3H8

seaquist@astro.utoronto.ca, swlee@astro.utoronto.ca

AND

G. H. MORIARTY- SCHIEVEN

Joint Astronomy Centre, University Park, Hilo HI 96720-6030

g.schieven@jach.hawaii.edu

## ABSTRACT

We present a map of $^{12}$CO J = 6-5 emission of the nuclear region of the nearby starburst galaxy M82 at resolution 7" taken with the James Clerk Maxwell Telescope (JCMT). This is the highest resolution map yet available at this transition. A detailed quantitative comparison is made with emission at $^{12}$CO J = 1-0 at the same resolution yielding new insights into the excitation of molecular gas in this galaxy. The excitation is found to be highest in the central area of the starburst region where the ratio $r_{61}$ = $^{12}$CO J = 6-5/$^{12}$CO J = 1-0 is as high as 0.5, compared to the mean value over the starburst region of 0.24. The excitation ratio peaks along the inner edge of the molecular ring outlined by atomic and molecular gas at lower excitation, and also in two spurs extending northward from the disk toward the outflow associated with the superwind. Emission with higher than average excitation is also found to be associated with the supershell surrounding the luminous SNR candidate 41.9+58, and possibly on a larger scale in gas whose orbits are strongly influenced by the stellar bar. The higher excitation in M82 is likely to be caused predominantly by local increases in kinetic temperature and/or in the geometric filling factor of a pre-existing higher excitation component, and less likely to be caused by local increases in gas density.

Subject headings: galaxies: individual (M82) – galaxies: ISM – galaxies: starburst - submillimeter

## 1. INTRODUCTION

The nearby dwarf starburst galaxy M82, at a distance of 3.25 Mpc (Sandage & Tammann 1975), is the most well studied starburst system, primarily because of the linear resolution afforded by its proximity. It is thus an ideal object for investigating rapid star formation, thought to be triggered in this case by a gravitational encounter with M81. The most recent encounter some $10^8$ years ago appears to have triggered a series of starbursts which are still continuing. Studies of the molecular component of the interstellar medium (ISM) are important in this context because it is this component that comprises the bulk of the gas, and it represents the cool dense gas from which stars form. Consequently, there are extensive studies of the ISM in the mm/sub-mm transitions associated with several molecules, each sensitive to different regimes of density and temperature. For example, $^{12}$CO has been studied in transitions ranging from J = 1-0 through J = 7-6, and various simple one or two component physical models developed using radiative transfer methods, most notably using the Large Velocity Gradient (LVG) approximation. Studies of these and other molecular transitions have determined that the molecular gas in star forming regions comprises at least two components, one diffuse and the other highly clumped, with densities spanning the range $10^3 - 10^5$ cm$^{-3}$ and kinetic temperatures 20 – 200K, and with a total mass of several times $10^8$ M$\odot$. There is also a powerful superwind or outflow containing comparable molecular mass driven by star formation with rates recently estimated to be 13 - 33 M$\odot$ yr$^{-1}$ (Förster Schreiber et al. 2003). Evidence for this star formation may be seen in phenomena across the entire spectrum, and is most directly evident in hundreds of super star clusters with masses $10^4$-$10^6$ M$\odot$ (Melo et al. 2005).

Success in modeling the molecular component of the ISM in M82 is limited by technical problems associated with mm and sub-mm observations, notably the lack of observations with common angular resolution at the level of few arcseconds (~ 100 pc at D=3.25 Mpc). In the case of $^{12}$CO J = 1-0 at 115 GHz, observations with angular resolution of a few arcseconds have been made with interferometers which often insufficiently sample the emission at low spatial frequencies. Observations at higher frequency suitable for comparison with the J = 1-0 transition have been made with single dish telescopes which sample the low spatial frequencies but provide lower angular resolution, typically 15-20 arcseconds (~ 300pc) or worse. In addition, single dish sub-mm observations are usually affected by relatively larger pointing and calibration errors. Thus reliable detailed comparisons covering a wide range of frequencies are difficult.

The work reported here represents a step forward by providing new observations of the $^{12}$CO J = 6-5 transition (691 GHz) at 7", the highest resolution to date at this frequency. Furthermore, these observations are compared quantitatively with high quality interferometric observations of $^{12}$CO J = 1-0 at 115 GHz by Walter et al. (2002) at a resolution of 3.6". The latter observations include single dish observations to fill in the low spatial frequencies not provided by the interferometer. These data can thus be smoothed to a spatial resolution of 7" for a reliable quantitative comparison with the $^{12}$CO J = 6-5 data.

The energy levels of the upper states of the $^{12}$CO J = 1-0 and $^{12}$CO J = 6-5 transitions are 5.5 K and 116 K respectively, providing a comparison between very different levels of excitation at 100 pc resolution. Though this study alone is insufficient for a full radiative transfer analysis, the resolution, range in excitation, and the high quality of the two data sets provide new insights into the relationship between gas conditions and star formation in the starburst region.

In section 2 we describe the observations. The results and analysis are described in section 3, and include a detailed comparison with the $^{12}$CO J = 1-0 data of Walter et al. (2002). Finally, we present in section 4 a detailed discussion of the significance of the results in terms of the star formation environment in the central 1 kpc of M82.

## 2. OBSERVATIONS

The observations were made with Receiver W on the James Clerk Maxwell Telescope (JCMT) in the $^{12}$CO J = 6-5 transition (691.473 GHz) during the period 2001 November 13-15. The observations were made in single sideband mode with the mixer tuned to the upper sideband. Although the receiver has two polarization channels, only a single channel was used to achieve sufficient spectrometer bandwidth to accommodate the velocity range in M82 at this high frequency. The autocorrelation spectrometer was operated in wideband mode offering a total bandwidth of 1.8 GHz (800 km s$^{-1}$ at 691 GHz) and channel separation 1.25 MHz (0.54 km s$^{-1}$). However, the data were binned to 20 spectral channels providing a resolution of 10.8 km s$^{-1}$. The weather was generally excellent, and the system temperature varied from about 5000 K to as high as 11000 K, primarily dependent on source elevation, but was typically near 5500 K. The observations were conducted in beam switching mode using a switch rate of 1 Hz and a beam throw of 120" in azimuth. The beamwidth (FWHM) of the JCMT was 7", and the main beam efficiency of the telescope was determined from two observations of Saturn to be 0.38. This value was found to be in agreement with a similar determination in 1999. This value for the main beam efficiency was used to convert the temperature scale from $T_A^*$ to $T_{mb}$, i.e. from antenna temperature to main beam brightness temperature. The uncertainty in this calibration is estimated to be about 15 percent.

Special observing procedures were adopted to measure the pointing corrections, since there were no sufficiently bright pointing calibrators near the position M82 at 691 GHz. First of all, the continuum bolometer SCUBA at 850μm was used to determine the SCUBA azimuth and elevation collimation errors about once every hour on either of the extragalactic sources 0836+710 or 1044+790. Both sources are near M82, and provided a high signal to noise ratio with this receiver. Then using the (fixed) collimation offsets between Receiver W and SCUBA measured using Saturn, the pointing offsets were determined for Receiver W and applied to the observations of M82. The collimation offsets between Receiver W and SCUBA were re-determined periodically during the run to ensure that they were constant. In this way, the pointing offsets for the M82 observations could be routinely and reliably determined from the two nearby calibrators. The scatter in the pointing offset measurements with Receiver W provided an estimate of the pointing uncertainty of 2.0".

Secondly, the M82 observations were made by combining many rapid and fully sampled grid-maps of the starburst region of M82 measuring 990 x 330 pc. The maps were aligned with the major axis of the starburst region (P.A. = 75°), comprising a rectangular grid of 19 x 7 or 133 individual points at intervals of 3.5" (ie half beamwidth). The center of the map grid and reference position for all images in this paper is $\alpha(1950) = 09^h 51^m 43.51^s$, $\delta(1950) = +69° 55' 01"$, essentially coincident with the 2μm nucleus (Dietz et al. 1986). The central and reference velocity is $V_{LSR} = 203$ km s$^{-1}$. The integration time on each grid point was 10 seconds per map, and a total of 12 such maps were co-added providing a total integration of 2 min per sampling point. The total time to produce one complete map was about one hour including overheads for receiver calibration and telescope moves. This procedure was adopted to minimize pixel registration errors in the map due to drift in the telescope pointing. The pointing observations on the two nearby standard sources described above were made between individual grid maps. The position registration between grid maps was checked by performing cross-correlations between one map and all other maps. The deviations in the locations the peaks of the cross-correlation functions yielded an rms registration error of 2.3" for the grid maps relative to one another, consistent with the estimated pointing errors (2.0") derived from the pointing source measurements.

3. DATA ANALYSIS AND RESULTS

The $^{12}$CO J = 6-5 data at 691 GHz are used together with data on $^{12}$CO J = 1-0 at 115 GHz from combined OVRO interferometer and IRAM 30m antenna observations (Walter et al. 2002), kindly provided for this comparison by F. Walter (private communication). The J = 1-0 data cover a much larger field of view than the J = 6-5 data, including the extended outflow, and also possess a higher spatial resolution (3.6"). Our comparison concerns only a limited part of the J = 1-0 map confined to the central 1 kpc region. For this purpose, the J = 1-0 data cube was smoothed to a spatial resolution of 7" and velocity channel separation of 10.8 km s$^{-1}$ to correspond to that of the J = 6-5 data.

3.1 Comparison of line intensities

The comparisons between the $^{12}$CO J = 6-5 and $^{12}$CO J = 1-0 data were made using the AIPS spectral line software package. Figures 1(a) and 1(b) show respectively spectral line profiles of the 133 points in our J = 6-5 grid map and corresponding points from the convolved J = 1-0 data cube. The correspondence between the two sets of profiles is generally good, but note that the main beam brightness temperatures and S/N ratios of the 691 GHz data are significantly lower than at 115 GHz. However it is the differences in distribution between the two sets of profiles that provide important insights into the excitation of the CO emitting gas, and which are the focus of this paper.

3.1.1 Integrated intensities

Figures 2(a) and 2(b) show contour maps at the same resolution of the intensities for J = 6-5 and J = 1-0 integrated over $V_{LSR}$ = 0 to 400 km s$^{-1}$. Both are overlaid with a grey scale image of the J = 1-0 integrated intensity image to facilitate comparison between the two transitions. The integrated flux (over area and velocity) in the J = 6-5 map is 2.1x10$^5$ K km s$^{-1}$ (arcsec)$^2$, somewhat less than in the recent lower resolution map of Ward et al. (2003) who obtain 2.8x10$^5$ K km s$^{-1}$ (arcsec)$^2$. The difference may be attributable in part to the different area covered and, partly due to errors in the calibration of the main beam efficiencies. The ratio of the integrated fluxes ($^{12}$CO J = 6-5/J = 1-0) over identical regions is 0.24, equivalent to the mean ratio weighted by the J = 1-0 brightness distribution.

A number of features of the J = 6-5 emission are evident from Figure 2. First, the brightest regions are displaced inward (toward the nucleus) with respect to the well known lobes at J = 1-0, usually designated as the molecular ring. The separation between the J = 6-5 peak intensities is about 17" (267 pc), whereas the separation of the CO (1-0) lobes is about 26" (408 pc). This displacement of excited CO has been previously noted by Mao et al. (2000), and is also seen in maps of other tracers of star formation, for example, the mid-IR associated with dust and [NeII] reflecting ionized gas (see table 2 of Mao et al. 2000 for an extensive tabulation of the lobe separation at different wavebands). There is also a peak near the radio luminous SNR candidate 41.9+58 and the molecular supershell centered on this source (Neininger et al 1998; Weiß et al. 2002; Matsushita et al. 2005). Third, prominent spurs of emission appear to project northward from the regions of prominent emission in the disk, at +9" and -3" (measured relative to the nucleus along the major axis). These spurs possibly reflect material associated with the molecular outflow in the superwind (see section 4.5).

3.1.2 Intensity ratios based on a regression analysis

Line intensity ratios $^{12}$CO J = 6-5/J = 1-0 (hereafter $r_{61}$) are traditionally computed using either the integrated or peak intensities. In this study, line ratios were obtained by performing a linear regression of our profiles against the corresponding J = 1-0 profiles at the same spatial and velocity resolution, with an allowance for any residual baseline offsets. For this purpose, the data cubes were sampled identically at half beam-width intervals and profiles extracted at identical velocities for the two transitions. Since the S/N ratio in the J = 1-0 profiles is much higher than for the J = 6-5 profiles, a standard linear least squares analysis was used with error assumed in the J = 6-5 coordinate only. If the profiles at both transitions are scaled versions of one another, the slope of the regression line yields the intensity ratio, and the standard error of the fit yields its uncertainty due to noise. In this case, the measure is equivalent to those based on either the integrated or peak intensities. If, however, corresponding profiles are not scaled versions of one another, then it may be shown that the regression method yields a value intermediate between the other two methods, and the standard error of the slope reflects both noise and the differences in profile shapes. This regression analysis provides a robust method of determining line ratios since it utilizes the entire line profile, is insensitive to the shape of the profile and to errors in the profile baselines (since these too are part of the fit).

The resulting line ratios and their uncertainties are plotted in Figure 3 as circles superposed on a grey scale map of the J = 1-0 integrated intensity. Figure 3 shows that the typical line ratio is about 0.2, but varies between 0.1 and 0.5. As indicated earlier, the mean ratio is 0.24. The lowest ratios are found at the peripheries of the starburst region, and the highest ratios are in the central regions, coincident with the locations of prominent features in the integrated J = 6-5 line emission, confirming that the latter are areas of higher than average molecular excitation. The ratios are highest ( up to 0.5) in the region of the aforementioned spurs at -3" (see Figure 2).

3.2 Gas Kinematics

The gas kinematics are explored through the use of channel maps and position-velocity plots (hereafter p-v plots) shown in Figures 4 - 8. All velocities given in the text and figures are relative to the reference velocity of the data cube at $V_{LSR}$ = 203 km s$^{-1}$.

3.2.1 Channel maps

Figure 4 shows a series of channel maps at J = 6-5 overlaid on the corresponding grey scale maps at J = 1-0 at the same resolution. Both data cubes were hanning smoothed to improve the S/N ratio of the J = 6-5 data. The resulting maps have a velocity resolution of 21.6 km s$^{-1}$. As is the case with the integrated emission, The channel maps in Figure 4 show that the brightest emission is displaced toward the inner part of the molecular ring outlined in the corresponding grey scale J = 1-0 channel images. In addition, there is a substantial northward displacement of the brightest J = 6-5 emission associated with the bright lobes, especially evident in channels with velocities -49 to -27 km s$^{-1}$ and +38 to +81 km s$^{-1}$. This displacement is unlikely to be associated with a pointing error since at other velocities the lobes at the two transitions seem well aligned. This displacement is associated with the spurs in the integrated intensity map of Figure 2, and appears to be excited gas entrained in the superwind. The emission may be associated with the base of the outflow, where superheated gas associated with supernovae is accelerated outward along the cone of the flow confined by the molecular ring. The channel maps also show an extension of emission to the south, which appear to be in good agreement in velocity and relative intensity with the J = 1-0 emission (see also the line ratio map in Figure 3 which shows no significant enhancement in the line ratio in this region).

The channel maps also reveal some insight into the higher J = 6-5 integrated intensity associated with the SNR candidate 41.9+58 evident in Figure 2. At this location the J = 6-5 emission covers a wider range in velocity than would be expected from the rotation curve (e.g. velocities in the range -135 to +38 km s$^{-1}$). A separate series of channel maps was also made from a data cube representing the difference or residual between J = 6-5 and J = 1-0, the latter scaled by a factor of 0.19. The latter factor is the average value of $r_{61}$ outside a radius of four sampling points (14") from the nucleus, representing therefore the regions of lowest excitation. Thus the computed residual cube reflects preferentially highly excited CO. These channel maps, shown in Figure 5, indicate excited CO overlapping with the SNR candidate at velocity channels -92 km s$^{-1}$ and -5 to +16 km s$^{-1}$. This emission is probably associated with the supershell centered on the SNR and the velocity range of this emission suggests an expansion rate of about 50 km s$^{-1}$ in good agreement with previous estimates.

3.2.2 Position-velocity plots

Figure 6 is a p-v contour plot of J = 6-5 emission along the major axis overlaid with a corresponding grey scale image at J = 1-0. Both contour and underlying grey scale images were made by averaging three contiguous p-v plots centered on the major axis to improve the S/N ratio at J = 6-5. The averaging thus covers an extent of 7" along the minor axis. The main feature reflects the rotation curve investigated by many previous authors. Figure 6 was used to obtain an estimate for the systemic velocity of $V_{LSR}$ = 213 ± 5 km s$^{-1}$ at J = 6-5 (or +10 km s$^{-1}$ relative to the reference velocity). Within the uncertainty, the dynamical center, i.e. the position on the major axis with this value, is coincident with the IR nucleus. As in previous images, the inward displacement by about 5" (79 pc) of the brightest parts of the molecular ring at J =6-5 is clearly evident. Also the emission at the position of the SNR candidate at -9" is extended in velocity relative to the rotation curve, both to lower and higher velocities. Emission occurs near -100 km s$^{-1}$, and also in a faint spur extending to about +50 km s$^{-1}$ (ie +100 km s$^{-1}$ relative to the rotation curve). The latter feature appears to be produced by excited gas associated with the receding part of the shell. This feature is not evident in the J = 1-0 emission. These structures are the counterparts of the supershell emission noted in the channel maps in Figures 4 and 5.

An interesting feature of the symmetry of the faint outer contours in the p-v plot in Figure 6 is an apparent faint extension of the J = 6-5 emission to high velocities on the east side and a similar extension to low velocities on the west side, all relative to J = 1-0. Its large scale and symmetry appear to rule out an association with the supershell centered on 41.9+58. This feature was examined further using a p-v plot representing of the excited component of the emission, shown in the form of channel maps in Figure 5. The p-v plot is shown in Figure 7, again averaged over three cuts parallel to and centered on the major axis, superposed on a J = 1-0 grey scale image. This plot of the excited emission confirms the faint high and low velocity extensions of excited CO emission relative to J =1-0 on the east and west sides of the nucleus respectively. Aside from an overlap with the supershell emission at -9", the feature covers the region between the brightest J = 6-5 emission at the inner edges of the main CO J =1-0 lobes. These features are difficult to produce by an instrumental effect, and we are inclined to accept it as real, albeit with caution. Figure 8 shows the corresponding p-v plot of the velocity dependent ratio $r_{61}$, where the plotted region has been clipped at $|T_{mb}| > 1.0$ K at J = 6-5 to ensure adequate S/N (>4) at both transitions. Not surprisingly, the intensity ratios also reflect the symmetric appearance of excited emission in Figure 7 and show that the peak line ratio $r_{61} > 0.7$ at the highest velocities at about +5" side and $r_{61} > 0.5$ centered on the IR nucleus.

4. DISCUSSION

The features described in section 3 have a strong bearing on the relationship between gas excitation and the star formation environment in M82, which we discuss below.

4.1 Relationship to star formation

The comparative distributions of J = 6-5 and J = 1-0 emission are broadly consistent with the description by Mao et al. (2000) that the primary lobes in the sub-mm and mid-far IR have a separation of about 15", considerably less than the 26" separation seen in the CO transitions J = 1-0 and J = 2-1 and in HI. Mao et al. suggest that much of this effect is due to broader line profiles in the higher excitation CO lines. Our results indicate some agreement with this interpretation since the p-v plot in Figure 6 shows J = 6-5 emission at -100 km s$^{-1}$ and 0 to +50 km s$^{-1}$ located at -9" (i.e. at the position of the SNR supershell) which is at the inner edge of the main J = 1-0 lobe. This broad velocity emission is not evident at J = 1-0. However, there are also peaks in J = 6-5 emission along the main ridge which are also displaced inward from the J = 1-0 lobes, showing that the effect is not solely attributable to broader line profiles. The inference is that J = 6-5 emission reflects the higher excitation of CO in the region where other star forming indicators are prominent (see below). Since there is little evidence that high density star formation tracers CS, HCN and HCO$^+$ are enhanced along this inner edge (Mao et al. 2000), a further inference is that the excitation of CO is primarily due to increased kinetic temperature or covering factor of excited gas the region, possibly associated with excess heating by UV from newly born stars (see also sections 4.2 and 4.5).

Figure 2(a) shows that there are several peaks in the J = 6-5 integrated brightness, with the outermost broad features coinciding with the primary J = 1-0 lobes shown in grey scale. The figure also shows the centers of the main [NeII] features E1,W1 and W2 noted by Achtermann & Lacy (1995). While there is no correspondence with E1 on the east side, the two [NeII] features on the west side lie within a broad prominent excess in CO J = 6-5 at the inner edge of the CO J = 1-0 lobes. The [NeII] feature at -6", designated W1 by Achtermann & Lacy, lies close to a peak in CO J =6-5, and both are nearly coincident with a CO J =1-0 peak termed "the central source" or C1 by Shen & Lo (1995). This broad feature in J = 6-5 also shows a peak at -10" coinciding closely with 41.9+58, and feature W2 in [NeII], slightly to the west of 41.9+58. Thus there appears to be a relationship between excited CO emission and features associated with star formation, but the relationship is complex and possibly strongly affected by emission from the supershell to the west of the nucleus, and possibly by molecular cloud extinction in the mid-IR on both sides.

4.2 The supershell associated with the prominent SNR candidate 41.9+58

As noted in previous discussion, evidence has been presented for the presence of a supershell seen in CO (Neininger et al. 1998; Weiß et al. 1999), the mm continuum (Matsushita et al. 2005).

It is also seen in absorption against the low frequency radio continuum (Wills et al. 1997). The shell has a diameter of about 130 pc and an expansion velocity of about 45 km s$^{-1}$ (Weiß et al. 1999). This may be the largest of a number of such shells seen in HI and distributed throughout the star forming region (Wills, Pedlar & Muxlow 2002; Pedlar, Muxlow & Wills 2003). Though its association with 41.9+58 suggests a SNe origin, the energy requirements are too large for a single SNe, and a more plausible explanation is formation by stellar winds and SNe from a massive cluster coinciding approximately with the luminous SNR candidate 41.9+58 (e.g. Yao et al. 2005). The kinematic evidence for the supershell appears most readily in CO p-v plots as a depression on the west side of M82, bounded by a feature emerging toward lower velocities and possibly blended with emission associated with gas following orbits in the bar potential. Neininger et al. (1998) conclude that depressions in p-v plots of $^{13}$CO J = 1-0 coincide with peaks in emission of [NeII] and radio recombination lines, providing evidence that the void is populated by ionized gas inside the supershell. Our p-v plot in Figure 6 reveals no depression in J = 6-5, but instead a region filled in with J = 6-5 emission which is not evident in the underlying J =1-0 map. The peak emission is at a velocity of about -110 km s$^{-1}$ and coincides with the location of the SNR, close to the western endpoint of the ring of ionized gas (source W2) suggested by Achtermann & Lacy (1995). The line ratio p-v map of Figure 8 shows a local peak value for $r_{61}$ of 0.3 at this location, significantly in excess of the typical line ratios in this plot. The intensity ratio based on the entire line profile at position -9" also shows a local peak at 0.28 ± 0.02 (see Figure 3). This is also consistent with the appearance of the channel maps which show emission in the shell region extending over a very broad range in velocity.

We conclude that the location of the supershell contains CO with higher than average excitation, together with the ionized gas. We also note that the hole created by the supershell is not associated with prominent emission in higher density tracers such as HCN and HCO$^+$ in their low excitation lines (e.g. Brouillet & Schilke 1993; Seaquist, Frayer & Bell 1998), indicating that the higher state of excitation may be due to higher kinetic temperature, as noted in section 4.1 for the star forming regions.

4.3 The spurs

The two northward extending spurs evident in Figure 2(a) appear to have their base in the middle of the disk region. In the channel maps of Figure 4, the western spur is evident over a channel range -92 to +38 km s$^{-1}$ and the eastern spur over a range +38 to +146 km s$^{-1}$, suggesting that this material is associated with the northern outflow from the starburst region. The eastern spur is coincident with the most prominent chimney noted by Wills et al. (1999), evident as a vertically structured void emerging northward from the disk. This void is attributed to absorption by ionized gas flowing into the halo. It is also coincident in the disk with the base of a prominent spur extending northward from the disk in the SiO v = 2-1 map of García-Burillo et al. (2001). Their interpretation is that this feature, which extends to a vertical height of 500 pc, is caused by shocks, likely a consequence of interaction between the outflow and entrained gas.

The western spur seems closely associated with the [NeII] feature W1 in the map of Achtermann & Lacy (1995) (also the central source C1 in the CO J = 1-0 map of Shen & Lo (1995) and evident in the underlying J = 1-0 map of Figure 2). It has been variously interpreted as an inner spiral feature (Shen & Lo 1995), the western edge of an ionized ring (Achtermann & Lacy 1995), and as part of the supershell associated with the SNR candidate 41.9+58 (Matsushita et al. 2005). It also coincides with a less prominent chimney in the 408 MHz map of Wills et al.

(1999). There appears to be no SiO feature associated with this spur although there is intense SiO emission associated with the nearby supershell. Regardless of the precise nature of the western J = 6-5 spur, it appears to be related to star formation activity in this region. As noted in section 3.1.2, the regression-based line ratios $r_{61}$ in this area reach values as high as 0.5, the highest anywhere in our map (see Figure 3).

4.4 Signature of the bar potential

The NIR reveals the existence of a nuclear bar with extent ~ 1 kpc along its major axis (Telesco et al. 1991). Nuclear bars have an important influence on the orbits of gas in the nuclear region. The observed consequences on gas motions in M82 have been discussed by Achtermann & Lacy (1995) for [NeII] and Neininger et al. (1998) for $^{13}$CO J = 1-0. Extending inward from the co-rotation radius at 500 pc, and terminating approximately at the Inner Lindblad Resonance (ILR), the gas follows stable, non-intersecting elliptical orbits whose major axes coincide with the axis of the bar (the x1 orbits). In the vicinity of the ILR these orbits become self intersecting, resulting in collisions between gas clouds, and the resulting dissipation causes the gas to settle into stable orbits whose axes are perpendicular to the axis of the bar (the x2 orbits). The x1 orbits lie along the rotation curve and the x2 orbits appear inclined with respect to the rotation curve in the p-v plane. The higher inclination of the x2 orbits is associated with their deeper location in the potential well of the bar and the close alignment between their axes and the line of sight (see Achtermann & Lacy 1995 and Neininger et al (1998) for plots showing the orbits).

As shown in Figures 6 - 8, and noted in section 3.2.2, there is weak excess J = 6-5 emission on both sides of the dynamical center associated with an axis inclined to the rotation curve in the p-v plane. This emission lies coincident with corresponding [NeII] and $^{13}$CO J = 1-0 located in x2 orbits extending to ± 6" (94) pc from the dynamical center. The maximum and minimum velocities reached at the eastern and western endpoints in Figure 6 are +150 km s$^{-1}$ and -110 km s$^{-1}$ respectively, closely similar to that for model x2 orbits in Neininger et al. (1998), which are about 175 km s$^{-1}$ and -120 km s$^{-1}$. Thus these regions of the p-v plane seem to have an excessive ratio $r_{61}$ and we tentatively suggest that preferentially excited gas is moving in x2 orbits. The enhanced excitation could be related to gas cloud collisions or star formation associated with the transfer of gas from x1 to x2 orbits.

4.5 Modeling the variation in the line ratios

The two high resolution data sets compared in this paper are insufficient for a full radiative transfer analysis. Such analyses for multi-transition CO lines at lower resolution (~20") and including high excitation transitions have been carried out by a number of authors (e.g. Güsten et al. 1993; Wild et al. 1992; Mao et al. 2000; Ward et al. 2003). Most of these studies have employed the LVG method of analysis. All studies find that a single gas component cannot adequately represent the data, and thus they usually incorporate two components - one diffuse with a high filling factor and the other in higher density clouds occupying a low filling factor. The derived parameters vary significantly from one model to another, an indication that the data are generally not sufficient in quality and quantity to permit a unique solution.

In this paper, we adopt one of these models as a lower excitation "baseline solution" for the outer part of the starburst region and explore how the parameters would have to be varied in order to account for the increases in $r_{61}$ seen in the central parts. Accordingly, we adopted a version of a model by Güsten et al. (1993), slightly modified to produce a lower mean line ratio $r_{61} \sim 0.2$ as seen in our results for the outer disk of M82. Based on an LVG solution, we lowered the density of the high temperature component (from $10^5$ cm$^{-3}$ to $10^4$ cm$^{-3}$) and obtained a satisfactory fit to all the line ratios reported in their paper together with $r_{61} = 0.2$. The adopted baseline conditions for the cool component are then T=25K, n=$10^3$ cm$^{-3}$ and for the warm component T=70K, n=$10^4$ cm$^{-3}$. The filling factors found by Güsten et al. for the two components are 0.26 and 0.08 respectively. In our modified model these values must be multiplied by a factor of about 4 to account for the higher brightness temperatures at J = 6-5 and J = 1-0 seen at our higher resolution. Thus the filling factor is near unity for the low temperature diffuse component. The increase in the filling factor with resolution implies that much of the structure in the lower resolution data possesses a spatial scale near the resolution of our data (~100 pc). Such dimensions would correspond to the major features of the CO distribution such as the molecular ring.

The observed increases in $r_{61}$ above the ambient value of 0.2, particularly at the inner edges of the main CO J =1-0 lobes and in the spurs, can be accommodated by a variety of conditions. We assume variations in the warm component only, since changing the cool component has minimal effect on increasing the intensity at J = 6-5. For example, an increment in density from $10^4$ cm$^{-3}$ to $3 \times 10^4$ cm$^{-3}$ at T=70K increases $r_{61}$ to 0.74, with little or no increase at still higher density. The limit is reached because a temperature of 70K is insufficient to permit significantly higher excitation. An increase in temperature from 70K to 130K at n=$10^4$ cm$^{-3}$ increases $r_{61}$ from 0.2 to 0.77, and up to unity at about 200K. The line ratio may also be increased to a limiting value of about 0.4 by increasing the filling factor of the warm component to 1.0. Among these possibilities, increases in $r_{61}$ are less likely attributable to density variations since, as noted in sections 4.1 and 4.2, there is little evidence for corresponding enhancements in high density tracers such as HCO$^+$ and HCN. Consequently the increased excitation is likely caused by an increase in the kinetic temperature by up to a factor of two, and/or a large increase in the filling factor of the warm component. Either condition would be consistent with increased volume density of stellar sources of UV in PDR regions, or by shocks. The cause of the excitation of CO in the x2 orbits of the bar is less clear since there has not been a search for gas in these orbits in higher density tracers.

## 5. CONCLUSIONS

The comparison between the distributions of $^{12}$CO J = 6-5 and $^{12}$CO J = 1-0 emission shows several notable features:

(1) The distribution of J = 6-5 emission is more concentrated toward the central area of the starburst region than is J = 1-0 emission. While the ratio $r_{61}$ obtained from the fluxes integrated over the entire starburst region is 0.24, the ratios at individual spatial points based on a comparison of line profiles varies from about 0.15 in the outer regions of the starburst to typically 0.25-0.30 in the central regions. The lack of corresponding enhancements in high density tracers such as HCN suggests that the higher line ratios are produced by enhancements in kinetic temperature or the filling factor of warm gas rather than in gas density.

(2) While the well known lobes associated with the ring in M82 are evident in J = 6-5, brighter emission is found to coincide with the star formation indicators such as [NeII] and mid-IR which are displaced inward toward the nucleus. On the west side there is a broad maximum in line ratio which coincides with the central source known as C1 in CO or W1 in [NeII]. This feature may be part of the ionized ring suggested by Achtermann & Lacy (1995) or an inner spiral feature. Part of this excited emission appears to be associated also with the supershell recently identified with the SNR 41.9+58. A kinematic study also confirms that enhanced J = 6-5 emission is associated with the supershell, filling the associated "hole" in $^{13}$CO (1-0) at velocities near -100 km s$^{-1}$. There is also evidence for enhanced emission in the range 0 to +50 km s$^{-1}$ representing the receding part of the shell. Both velocities are measured with respect to the reference velocity of 203 km s$^{-1}$.

(3) The J = 6-5 emission highlights two spur-like structures with a high ratio $r_{61}$ extending northward from regions of enhanced star formation in the disk. One of these on the east side of the nucleus coincides with the most prominent chimney seen in the continuum at low radio frequency, and with the base of a prominent spur seen in SiO extending several hundred pc northward from the disk. The other feature, on the west side of the nucleus, coincides with a less prominent chimney and with star forming indicators such as [NeII]. The spurs appear to represent shock heated molecular gas entrained by warm gas flowing out of the disk forming the superwind. It is not clear why no similar structures in excited CO appear on the southern side of the disk, since the outflow occurs both above and below the disk.

(4) The p-v plot along the major axis shows strong hints of excess J = 6-5 emission lying above (higher velocities) the rotation curve on the east side of the dynamical center, and below (lower velocities) the rotation curve on the west side. Such gas is seen also in [NeII] and $^{13}$CO J = 1-0 and may be produced by excited gas in bar-induced x2 orbits. The higher excitation may be caused by gas cloud collisions near the Inner Linblad Resonance where the x1 orbits intersect one another transferring gas to the x2 orbits.

We thank Fabian Walter for providing the $^{12}$CO J = 1-0 data in electronic format for comparison with our $^{12}$CO J = 6-5 data. We also thank Ms. Lihong Yao for her assistance with the map cross correlation analysis. This work was conducted with the assistance of a Discovery Grant to E.R.S. from the Natural and Engineering Research Council of Canada.

FIGURE CAPTIONS

Figure 1

(a) Top: Grid map of J = 6-5 profiles based on observations reported in this paper. Grid spacing is 3.5". The horizontal and vertical offsets refer to are respective parallel to the major and minor axes, and are measured with respect to the reference position $\alpha$ (1950) = $09^h\ 51^m\ 43.51^s$ and $\delta$ (1950) = +69° 55' 01". The same coordinate system and reference position applies to all figures in this paper.

(b) Bottom: Grid map of J = 1-0 profiles for comparison with J = 6-5 profiles, prepared from data published in Walter et al. (2002). These data have been smoothed to the spatial and velocity resolution of the J = 6-5 data (7" and 10.8 km s$^{-1}$ respectively) and sampled at the same points.

Figure 2

(a) Top: Contour map of J = 6-5 intensities integrated from $V_{LSR}$ = 0 to 400 km s$^{-1}$ shown superposed on a grey scale map of CO J = 1-0 integrated intensities based on data from Walter et al. (2002) convolved to the same resolution (see also Figure 2 (b)). Only data with $T_{mb}$ > 1.0 K (~2.5$\sigma$) are included in the integration. Contours are (-1, 1, 2, 3, 4, 5, 6) x 50 K km s$^{-1}$ and the grey scale intensity wedge in units of $10^3$ K km s$^{-1}$ is shown at the top. The cross refers to the position of the NIR nucleus measured by Dietz et al. (1986) and the star marks the position of the SNR candidate 41.9+58. The smaller symbols refer to features in the [NeII] map of Achtermann & Lacy (1995), namely E1 (square), W1 (triangle) and W2 (diamond). Feature W1 is also coincident with the central source, termed C1 by Shen & Lo (1995). The FWHM beamsize (7") is shown at lower right.

(b) Bottom: Contour and grey scale map of the J = 1-0 integrated intensities shown in Figure 2(a) to facilitate comparison with this figure. Contours are (-1, 1, 2, 3, 4, 5, 6) x 250 K km s$^{-1}$, and the grey scale intensity wedge in units of $10^3$ K km s$^{-1}$ is shown at the top. Symbols have same meaning as in Figure 2(a).

Figure 3

The distribution of line ratios $r_{61}$ at 3.5" grid intervals superposed on the grey scale version of the CO J = 1-0 integrated intensity map for comparison. See the text for details. The values of $r_{61}$ are represented by circles with sizes proportional to the magnitudes of the ratios, and this representation is restricted to cases where the ratio exceeds 3$\sigma$. Otherwise the 3$\sigma$ upper limit on the ratio is represented by a cross to the same scale. The scale for the ratios is displayed at the bottom of the figure representing ratios (from left to right) of 0.1, 0.2, 0.3, 0.4, and 0.5. The cross within the circle at the center of the map marks the location of the near IR nucleus.

Figure 4

Channel maps of J = 6-5 maps (contours) superposed on J = 1-0 channel maps (grey scale) at the same resolution and velocity resolution smoothed to 21 km s$^{-1}$. Contours are (-3, -2, -1, 1, 2, 3, 4, 5, 6) x 0.6 K and the pixel intensity scale for the grey scale map (in K) is shown in the wedge above. The figures in the upper right corner indicate the channel velocities in km s$^{-1}$ relative to the reference velocity $V_{LSR}$ = 203 km s$^{-1}$ (applicable also to all other figures). The symbols showing the nucleus and SNR candidate 41.9+58 are as in Figure 2(a). The FWHM beamsize (7") is shown in the lower right panel.

Figure 5

Same as Figure 4, but showing only contours from highly excited emission at J=6-5, obtained by subtracting the J=1-0 map scaled by 0.19 (see text for details). Contours are (-3, -2, -1, 1, 2, 3) x 0.6K, and symbols are the same as in Figure 4. Note that some negative contours may reflect low excitation emission rather than noise ($\sigma$ ~ 0.3K).

Figure 6

Position–velocity (p-v) plot of J = 6-5 emission (contours) along the major axis superposed on the corresponding J = 1-0 plot (grey scale). The plots are smoothed along the minor axis by averaging three individual p-v plots separated by 3.5" (HPBW) centered on the major axis. Contours are (-3, -2, -1, 1, 2, 3, 4, 5, 6, 7, 8) x 0.5 K and the grey scale intensity wedge in K is shown at the top. The arrow on the position axis refers to the location of the SNR candidate 41.9+58 (also for Figures 7 and 8).

Figure 7

Position-velocity (p-v) plot of excess J = 6-5 emission (contours) defined as in Figure 5, superposed on p-v plot of J = 1-0 emission (grey scale) and smoothed along the minor axis as in Figure 5. The plot emphasizes the distribution of highly excited CO by removing the low excitation CO characterized by $<r_{61}>$ = 0.19 in the outlying areas of the nuclear region. Contours are (-3,-2,-1, 1, 2, 3) x 0.5K. Note that some negative contours may reflect low excitation emission rather than noise ($\sigma$ ~ 0.25K).

Figure 8

Position-velocity (p-v) plot of the ratio $r_{61}$ restricted to $|T_{mb}|$ > 1.0 K (~4$\sigma$) with contours superposed on grey scale of the same image. Contours in $r_{61}$ are (-1, 1, 2, 3, 4, 5, 6, 7, 8, 9, 10) x 0.075, and the intensity wedge along the top shows the ratio x 1000. Smoothing of the p-v intensity plot data contributing to this ratio map are as described in Figures 6 and 7.

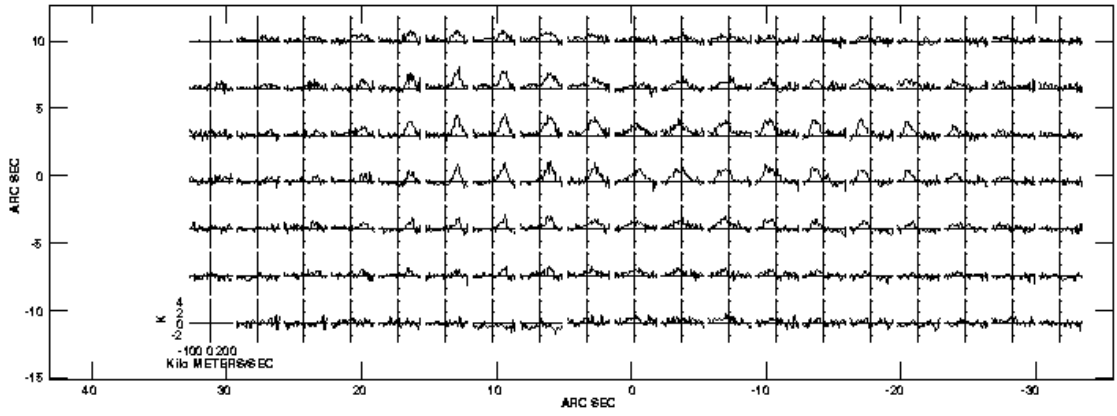

Figure 1(a)

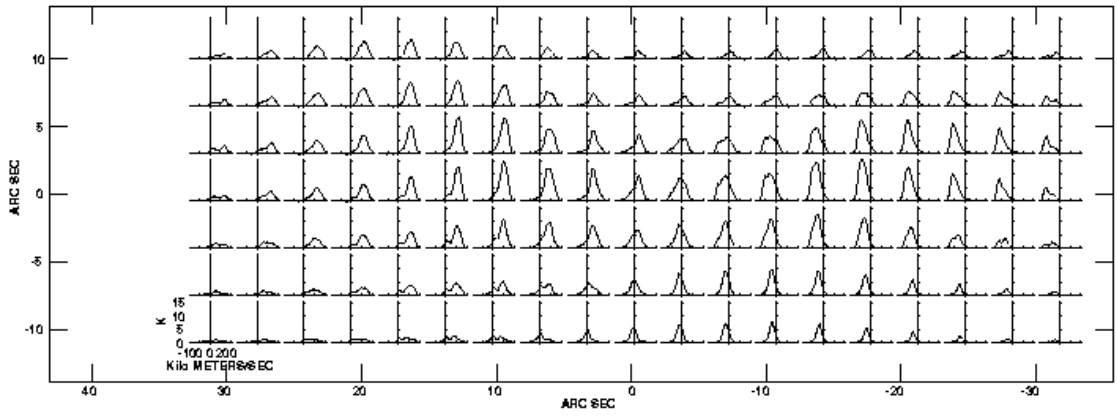

Figure 1(b)

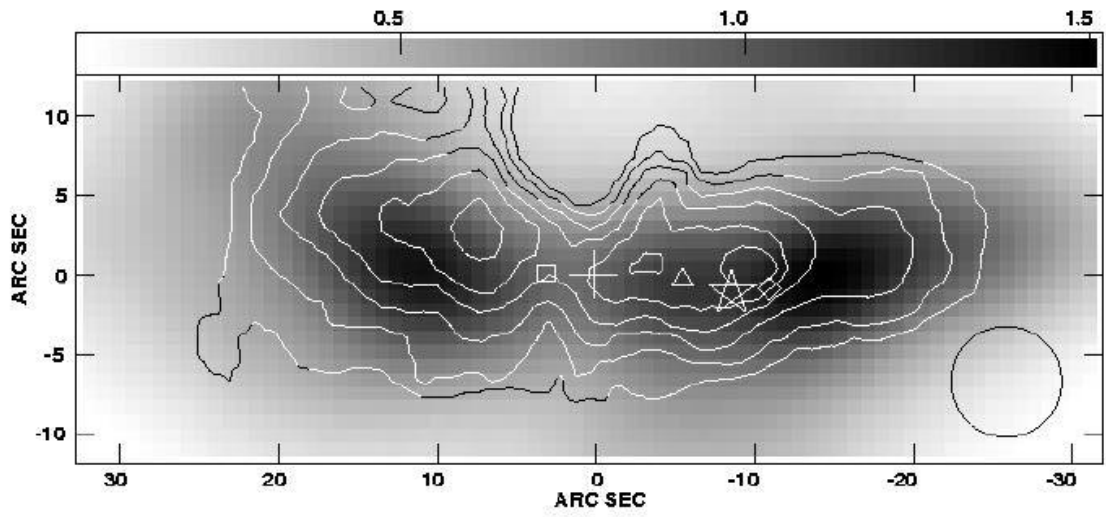

Figure 2(a)

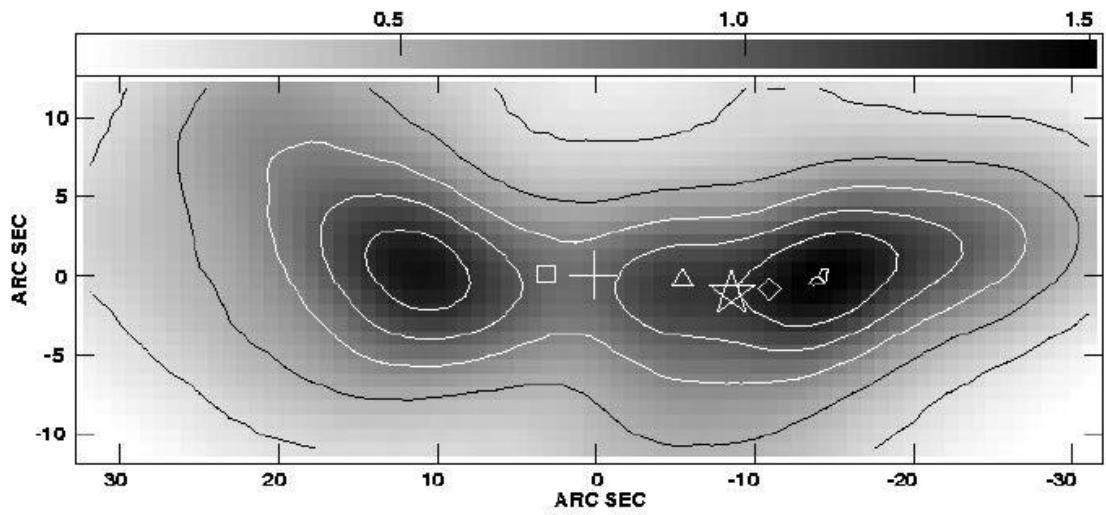

Figure 2(b)

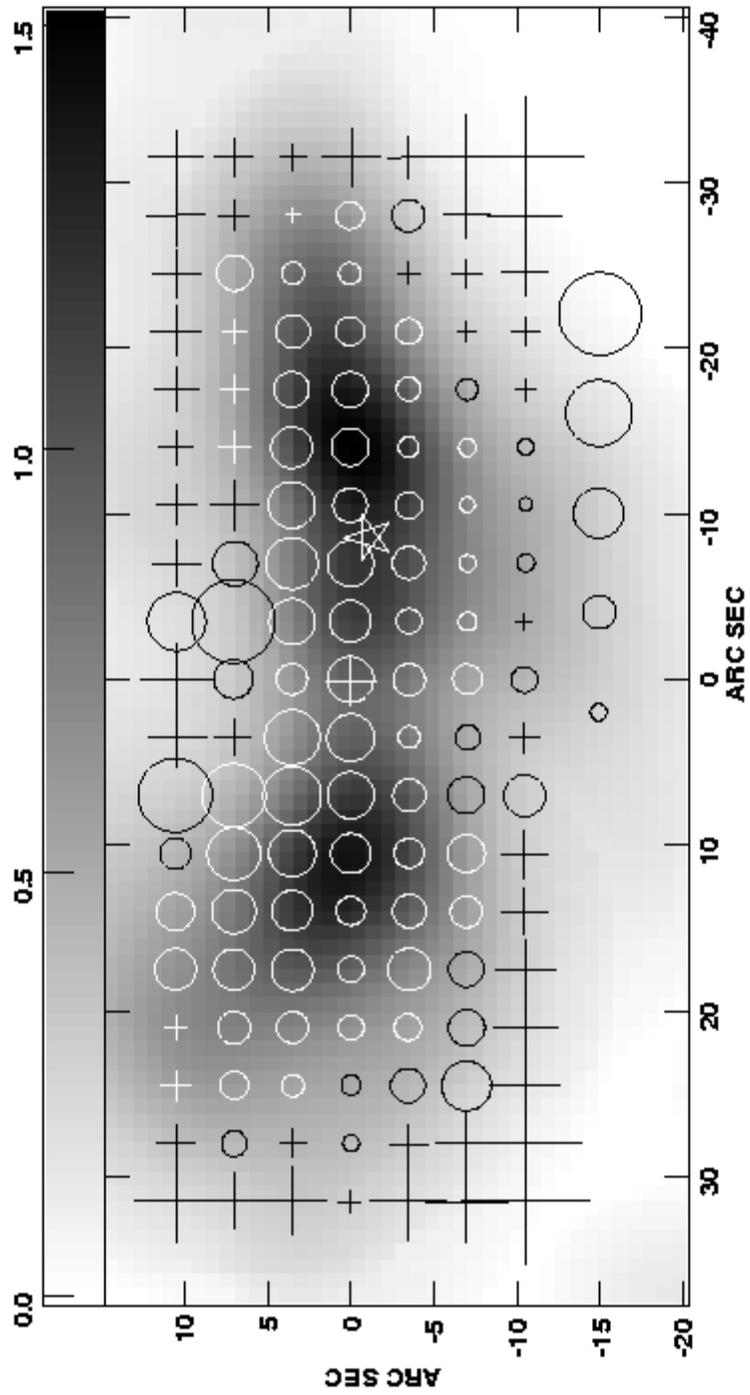

Figure 3

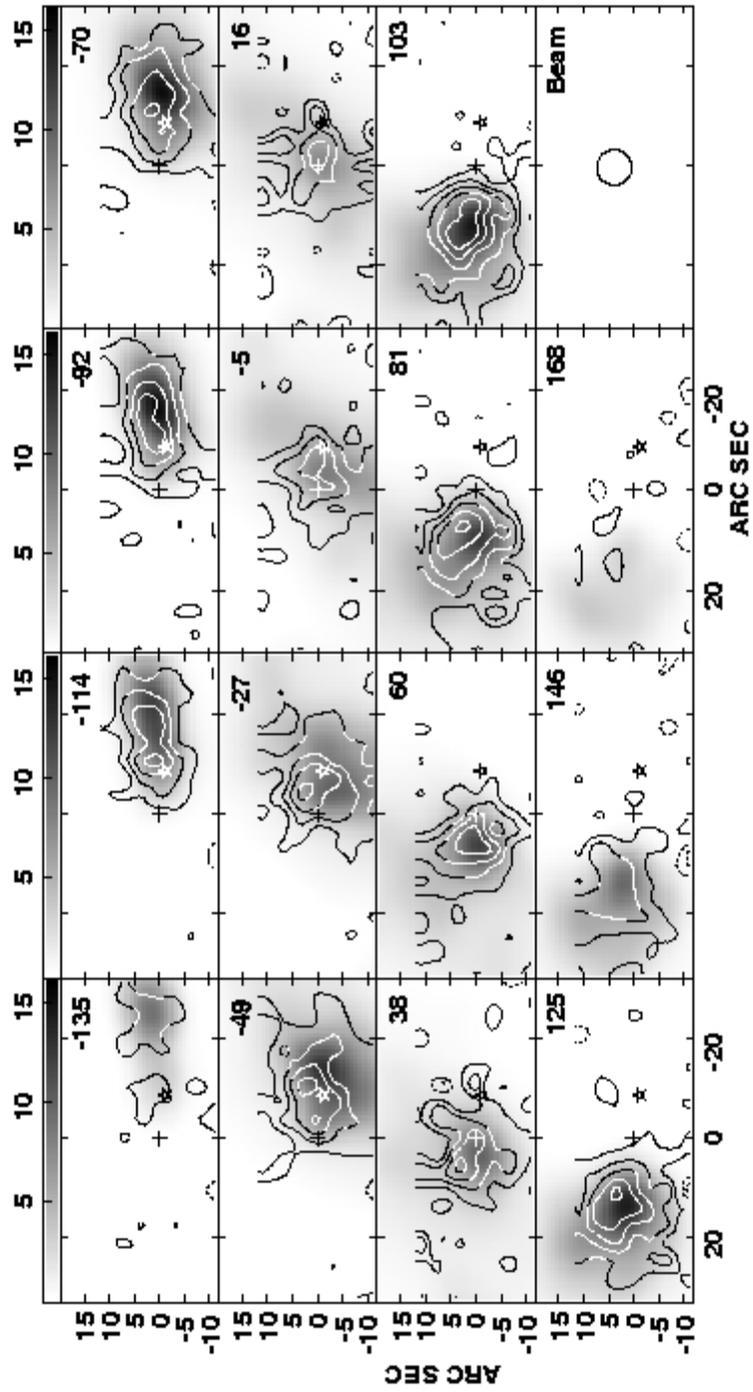

Figure 4

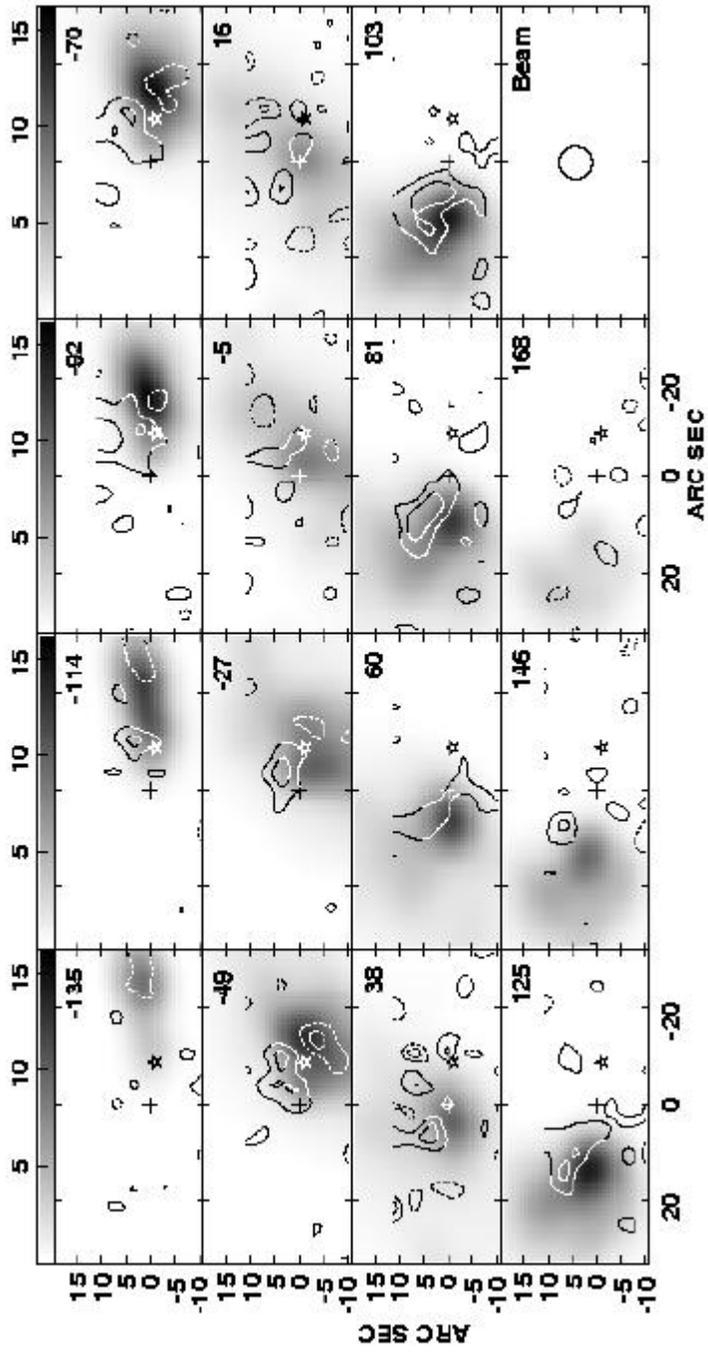

Figure 5

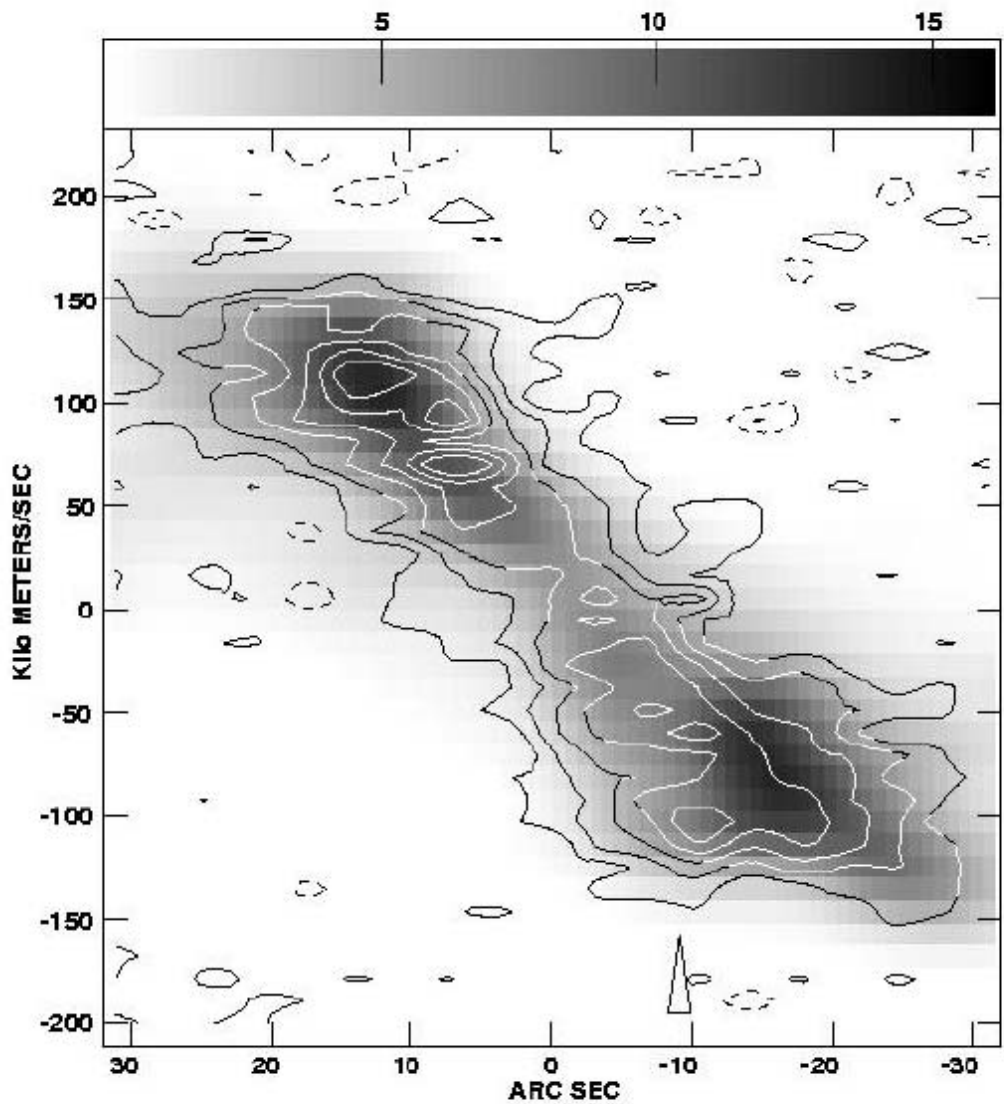

Figure 6

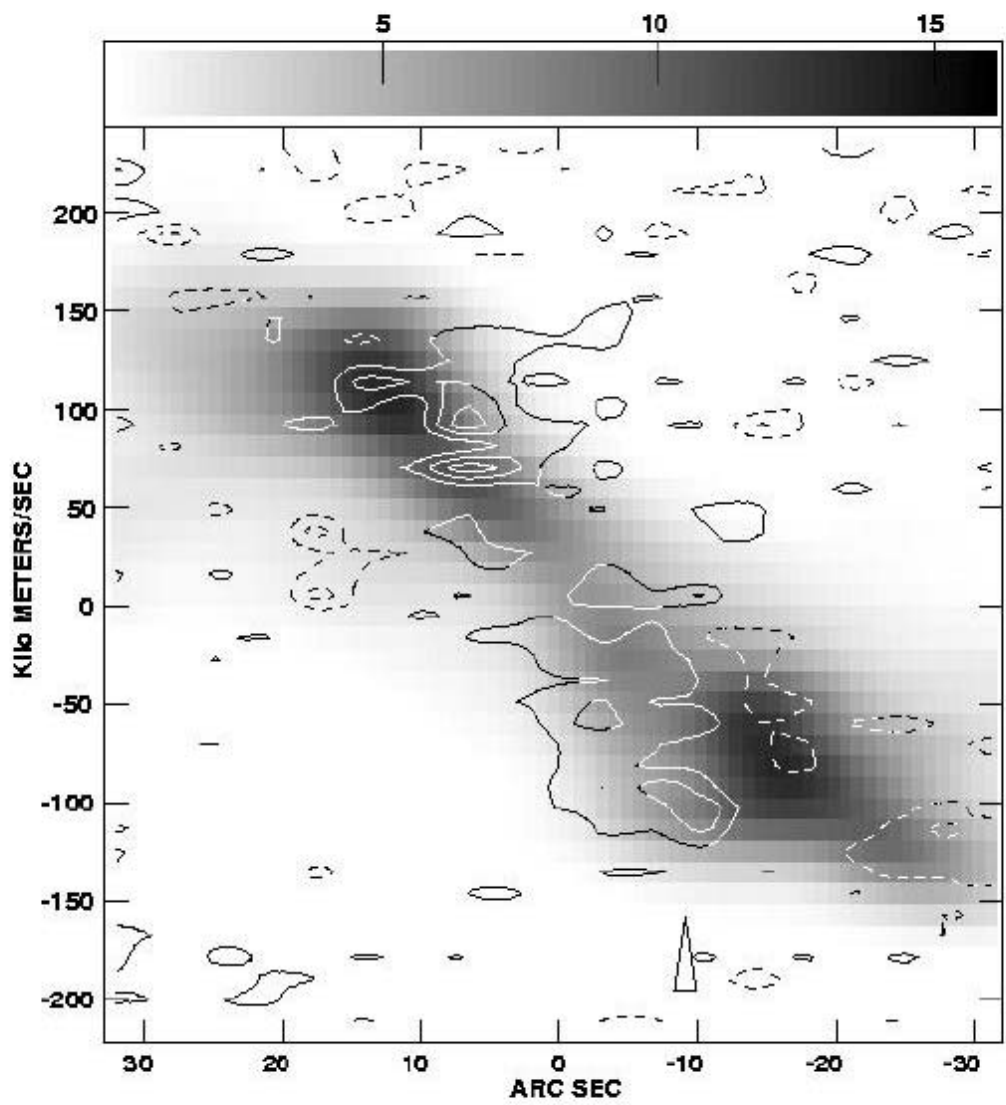

Figure 7

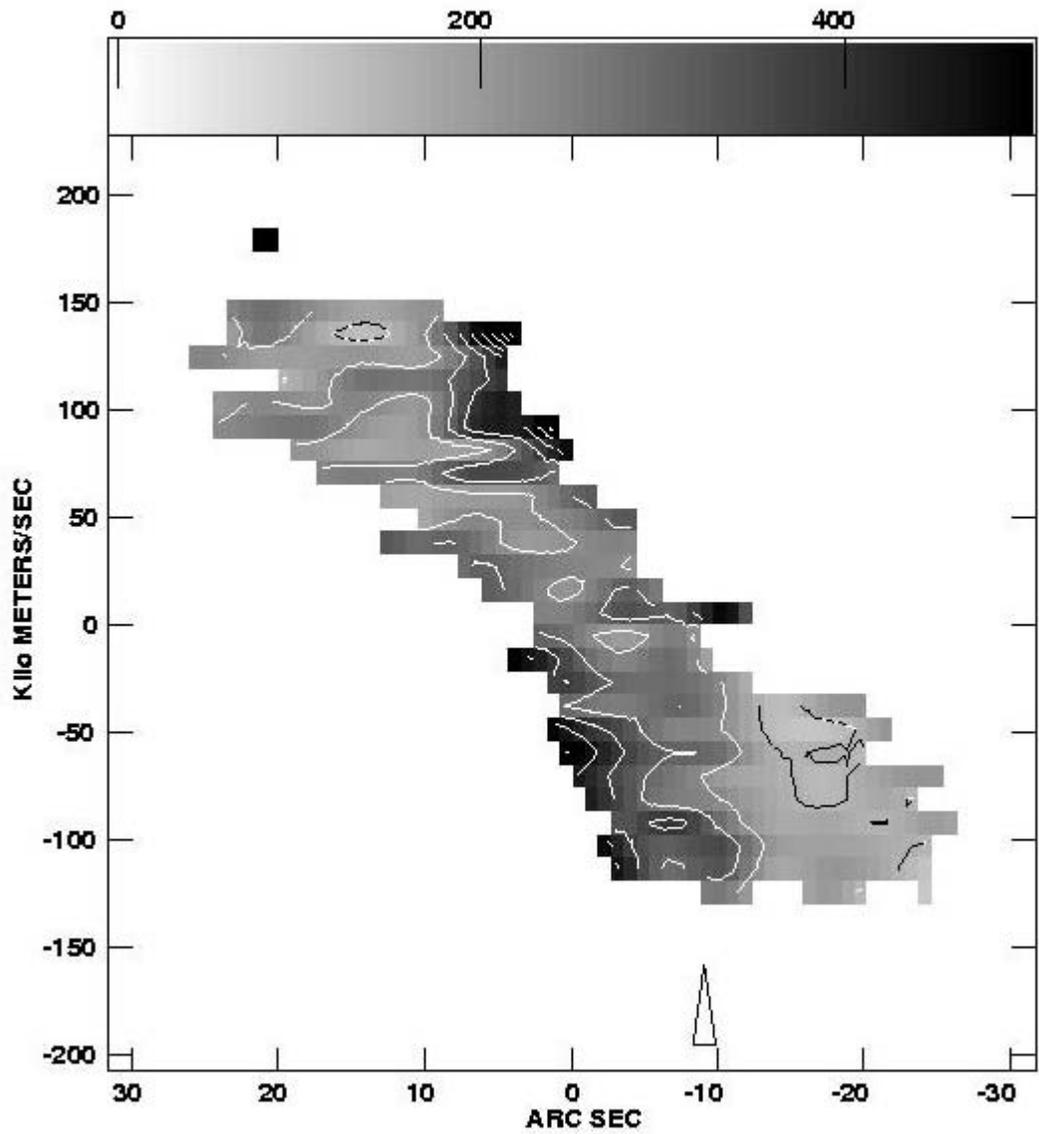

Figure 8